\documentclass[twoside,slac_one]{revtex4}%
\usepackage{fancyhdr}
\usepackage{amsmath} 
\usepackage{bm}
\usepackage{amsxtra}
\usepackage{amssymb}
\usepackage{amsthm}
\usepackage{latexsym}
\usepackage{lscape}
\newcommand{\ep}{\varepsilon}
\renewcommand{\vec}[1]{{\bm{#1}}}
\begin{document}
\title{The Epsilon Expansion of Feynman Diagrams via 
Hypergeometric Functions and Differential Reduction}
%
\author{S.A. Yost}
\affiliation{Department of Physics, The Citadel, Charleston, SC, USA}
\author{V.V. Bytev}
\affiliation{Joint Institute for Nuclear Research, Dubna, Russia}
\author{M.Yu. Kalmykov}
\author{B.A. Kniehl}
\affiliation{{\sl II} Institut f\"ur Theoretische Physik, 
Universit\"at Hamburg, Hamburg, Germany}
\author{B.F.L. Ward}
\affiliation{Department of Physics, Baylor University, Waco, TX, USA}

\begin{abstract}
Higher-order diagrams required for radiative corrections to mixed electroweak 
and QCD processes at the LHC and anticipated future colliders will require 
numerically stable representations of the associated Feynman diagrams. The 
hypergeometric representation supplies an analytic framework that is useful 
for deriving such stable representations. We discuss the reduction of Feynman 
diagrams to master integrals, and compare integration-by-parts methods to 
differential reduction of hypergeometric functions. We describe the problem 
of constructing higher-order terms in the epsilon expansion, and characterize 
the functions generated in such expansions.
\end{abstract}

\maketitle

\thispagestyle{fancy}

\section{Introduction\label{Intro}}

A variety of approaches are known for evaluating Feynman diagrams
analytically \cite{smirnov} and constructing the $\ep$ expansion of 
dimensionally-regularized integrals. \cite{dimreg}
One-loop diagrams \cite{1L,FJT,KT2009,KT2010}  
are expressible in terms of generalized 
hypergeometric functions of the form ${}_{p+1}F_{p}$, Appell functions 
$F_1$, $F_2$, $F_3$, $F_4$, or Horn-type \cite{GKZ} multivariable 
functions. \cite{FD} 
Analytical techniques for evaluating the finite part of one-loop diagrams 
have long been known, \cite{tHV} but higher order diagrams require more terms 
in the $\ep$ expansion.  These facts motivate us to seek a way to obtain 
the all-order $\ep$ expansion of Horn-type hypergeometric functions in general. 

The first such construction of the all-order $\ep$ expansion was obtained 
for the Gauss hypergeometric function with special, physically-motivated,
values of the parameters. \cite{davydychev} The first systematic algorithm
for constructing the $\ep$ expansion of multi-variable hypergeometric
functions was described in Ref.\ \cite{MUW}, using Goncharov's 
polylogarithms \cite{Goncharov} in the case with integer values of parameters.
However, diagrams with massive propagators lead to hypergeometric functions
with rational values of parameters. \cite{example2} This case generates
multiple (inverse) binomial sums which were first systematically investigated 
in Refs.\ \cite{KV00,JKV,DK04}. Ref.\ \cite{Weinzierl:2004} generalized the 
technique of Ref.\ \cite{MUW} to the case of rational parameters, but the 
method was limited to the ``zero-balanced'' case, and applied 
to ${}_{p+1}F_{p}$ and $F_1$, but not $F_2$ or $F_3$.  
Some statements about the structure of the coefficients
of the $\ep$-expansion for ${}_{p+1}F_{p}$ with one non-balanced 
rational parameter contradicted explicit calculations of the first 
few coefficients of the $\ep$ expansion. \cite{DK04,MYK04}
In particular, 
the multiple sum
$
\sum_{j=1}^\infty z^j [j\binom{2j}{j}]^{-1} \sum_{k=1}^{2n-1} k^{-2}
$
was evaluated analytically in Ref.\ \cite{DK04} (see Eq.\ (C.22) there),
and it was found that the $\ep$ expansion of the 
hypergeometric function
\begin{equation}
\hspace*{-3mm}
_{p+1}F_p\left(\begin{array}{c|}
I_1 +\tfrac{1}{2} + b_1 \ep \;, K_j+a_j \ep \\
I_2 +\tfrac{1}{2} + f_1 \ep,    I_3 +\tfrac{1}{2} + f_2 \ep, \; L_i+c_i\ep \; 
\end{array} ~z \right)
\label{contradiction1}
\end{equation}
with integer values of $I_j, K_j, L_j$ leads, at weight ${\bf 3}$, to 
classical polylogarithms of argument $\frac{1-\sqrt{y}}{1+\sqrt{y}}$. 
(This was later confirmed in Ref.\ \cite{Maitre}.)
Furthermore, Ref.\ \cite{MYK04} analyzed the more general case 
\begin{equation}
\hspace*{-3mm}
_{p+1}F_p\left(\begin{array}{c|}
I_1 + \tfrac{1}{2} + b_1 \ep \;,
I_2 + \tfrac{1}{2} + b_2 \ep \;,
K_j + a_j \ep \\
I_3 +\tfrac{1}{2} + f_1 \ep,    I_4 +\tfrac{1}{2} + f_2 \ep, I_5 
+\tfrac{1}{2} + f_3 \ep, \; L_i+c_i\ep \; 
\end{array} ~z \right),
\label{contradiction2}
\end{equation}
via the sum $ \sum_{j=1}^\infty z^j[ j\binom{2j}{j}]^{-1} 
\sum_{k=1}^{2n-1} k^{-3}$, and was again found to differ 
from the prediction of Ref.\ \cite{Weinzierl:2004}.

The inadequacies of existing algorithms and a desire to explore the 
results of Ref.\ \cite{FJT} for one-loop diagrams motivated the development
of new technologies \cite{KWY,YKW:F1,Quarks08,KK1,KK2,KK3} and the extension 
of previous ones \cite{DK04} for all-order expansions, \cite{series:all-order} 
with the goals of 
\begin{itemize}
\item[(i)] independently verifying the results of Ref.\ \cite{Weinzierl:2004},
\item[(ii)] constructing the analytical coefficients of the $\ep$ expansion of 
hypergeometric functions of several variables. 
\end{itemize}
Recently, new generalizations of the algorithm of Ref.\ \cite{MUW} have
been found using cyclotomic harmonic sums \cite{cyclo} and 
$Z$-sums, \cite{Z-sums} allowing the $\ep$ expansion to be constructed 
in the cases of Eqs.\ (\ref{contradiction1}) and (\ref{contradiction2}).
Their applicability to multivariable hypergeometric functions is not yet clear.

The higher terms in the $\ep$ expansion of one-loop diagrams can be 
constructed \cite{tHV} without applying algorithms for expanding 
hypergeometric functions. \cite{one-loop:linear,KMR,pentagon,hexagon}
However, results derived via the hypergeometric representation are more
compact and often can be expressed in terms of simpler functions,
{\it e.g.}\ Nielsen's polylogarithms instead of Goncharov's. This can be seen,
for example, by comparing the results presented in 
Refs.\ \cite{one-loop:linear,KMR} to those in 
Refs.\ \cite{davydychev,DK04,vertex-ep}.  Some remarkable 
results on hexagon diagrams were presented in Ref.\ \cite{hexagon}, but this 
was done via an experimental mathematical analysis \cite{exp} specialized 
to this case. Methods following from the properties of hypergeometric 
functions hold the promise of being applicable to wide classes of 
Feynman diagrams. In addition, the $\ep$ expansion of hypergeometric 
functions has interesting applications beyond Feynman diagrams, in a 
broader mathematical context. \cite{Borwein}

In this paper, we will 
concentrate on Horn-type hypergeometric functions, a very general class which
encompasses all of those mentioned earlier, and which we 
conjecture \cite{conjecture} to be general enough to express all Feynman 
diagrams.  Section \ref{Reduction} describes the differential reduction of 
this class of hypergeometric functions. 
Section \ref{Basis} describes the differential reduction 
in application to Feynman diagrams 
and explain the difference in counting master integrals 
via our differential technologies \cite{KK2011}
versus integration-by-parts techniques \cite{ibp}. Section \ref{Expansion}
describes our approach to constructing the $\ep$ expansion, and provides some
examples. 

\section{Differential Reduction of Horn-type Hypergeometric Functions
\label{Reduction}}

Let us begin by considering the generalized hypergeometric 
functions $_{p+1}F_{p}(\vec{a};\vec{b};z)$ defined by a series about $z=0$ as
\begin{eqnarray}
{}_{p+1}F_{p}(\vec{a};\vec{b};z) 
= \sum_{j=0}^\infty 
\frac{\Pi_{i=1}^{p+1}(a_i)_j}{\Pi_{k=1}^{p}(b_k)_j} \frac{z^j}{j!} \;, 
\label{definition}
\end{eqnarray}
where $(a)_k = \Gamma(a+k)/\Gamma(a)$ is the Pochhammer symbol, and
$\vec{a}=(a_1,\cdots, a_p)$, $\vec{b}=(b_1,\cdots, b_q)$ are
called the upper and lower parameters, respectively. This function satisfies
a differential equation of order $p+1$, so that there are $p+1$ independent 
solutions for unexceptional values of the parameters. 

``Differential reduction'' is an algorithm by which any 
function ${}_{p+1}F_{p}(\vec{a}+\vec{m};\vec{b}+\vec{k}; z)$ with integer 
parameter lists $\vec{m}, \vec{k}$ may be expressed in terms of a set of 
$p$ functions in which the values of these arguments are shifted by 
lists of integers $(\vec{l})_j$ and $(\vec{r})_j$, respectively, 
with $j = 1, \ldots p$.  This reduction has the form 
\begin{eqnarray}
\label{decomposition:1a}
Q_{p+1}(\vec{a},\vec{b},z)
{\ }_{p+1}F_{p}(\vec{a}+\vec{m};\vec{b}+\vec{k}; z)
= 
\sum_{j=0}^{p}
Q_j(\vec{a},\vec{b},z) {\ }_{p+1}F_{p}(\vec{a}+\vec{l}_j;\vec{b}+\vec{r}_j; z) 
\end{eqnarray}
where the $Q_j$ are a set of polynomials in $a_i, b_i$, and $z$.
The resulting expression can be converted into derivatives of the unshifted
hypergeometric function, so that the reduction takes the form \cite{takayama}
\begin{eqnarray}
\label{decomposition:1b}
&& 
S(\vec{a},\vec{b},z) {\ }_{p+1}F_{p}(\vec{a}+\vec{m};\vec{b}+\vec{k}; z) = 
\sum_{j=0}^p R_j(\vec{a},\vec{b},z)\ \theta^j 
{\ }_{p+1}F_{p}(\vec{a};\vec{b}; z) \;,
\end{eqnarray}
where $\theta = z d/dz$ and $S$ and $R_i$ are other polynomials in 
$a_i, b_i$, and $z$. 

For particular values of the parameters, an algebraic solution of the 
corresponding differential equations can be found, so that the differential 
reduction simplifies.  For example, when some of the upper parameters are 
integers $I_i$, we have \cite{BKK2009}
\begin{eqnarray}
\widetilde{P}(\vec{a},\vec{b},z)
{\ }_{p+1}F_{p}(\vec{I}, \vec{a}+\vec{m};\vec{b}+\vec{k}; z)
= 
\sum_{j=0}^{p-1}
\widetilde{R}_{j}(\vec{a},\vec{b},z)\ \theta^j
{\ }_{p+1}F_{p}(\vec{I}, \vec{a};\vec{b}; z) 
+ 
\widetilde{R}_{p}(\vec{a},\vec{b},z)
\;,
\label{decomposition:1}
\end{eqnarray}
with algebraic functions $\widetilde{R}_{p}(\vec{a}, \vec{b}, z)$.
In this case, the Eq.~(\ref{decomposition:1a}) has the form
\begin{eqnarray}
\label{decomposition:1c}
\widetilde{Q}_{p+1}(\vec{a},\vec{b},z)
{\ }_{p+1}F_{p}(\vec{I},\vec{a}+\vec{m};\vec{b}+\vec{k}; z)
= 
\sum_{j=0}^{p-1}
\widetilde{Q}_j(\vec{a},\vec{b},z) 
{\ }_{p+1}F_{p}(\vec{J},\vec{a}+\vec{l}_j; \vec{b}+\vec{r}_j; z) 
+ \widetilde{R}(\vec{a}; \vec{b}; z)
\;,
\end{eqnarray}
where $\vec{I},\vec{J}$ are lists of integer parameters. 

This reduction can be generalized to all multivariable Horn-type 
hypergeometric functions \cite{takayama} having $L$ independent solutions:
\begin{itemize}
\item[(i)] For unexceptional values of parameters (irreducible monodromy group),
linear differential relations can be found among $L+1$ functions with parameters
differing by integers. 
\item[(ii)] When the monodromy group is reducible, the number of functions 
entering into these differential relations is correspondingly reduced.
\end{itemize}
In the reducible case, a basis can currently be constructed only when
the differential reduction has been carried out explicitly.

\section{Counting the Number of Basis Elements\label{Basis}}
\subsection{Preliminary Considerations\label{Preliminary}}
Consider a function $f(z)$ satisfying a homogeneous differential equation of
order $k$ with polynomial coefficients:
\begin{eqnarray}
&& \sum_{j=0}^k a_j(z) \left(\frac{d}{dz} \right)^j f(z) = 0 \;.
\end{eqnarray}
It is easily shown that the function $H(z) = z^a f(z)$ satisfies a 
differential equation of the same order as $f(z)$. 
Consider a function $H(z)$ which is a linear combination of functions 
$f_j(z)$ satisfying homogeneous differential equations of order 
$r_k$ with rational coefficients, 
\begin{eqnarray}
&&
H(z) = \sum_{k=1}^m c_k x^{\alpha_k} f_k(z) \;, 
\\ && 
\left(\frac{d}{dz} \right)^{r_k} f_k(z) 
+ 
\sum_{j=0}^{r_k-1} a_j(z) \left(\frac{d}{dz} \right)^j f_k(z) = 0 \;,
\end{eqnarray}
where $\alpha_k$ are rational numbers and $r_k$ are integers. 
Then it can be shown that the function $H(z)$ satisfies a differential 
equation with rational coefficients of order not less than $m \times r_k$.

We can take the $f_k$ to be Horn-type hypergeometric functions 
(multi-variable, in general)
whose parameters are linear combinations of the parameters of some
Horn-type function $H(\vec{J};\vec{z})$, which plays the role of $H(z)$ above. 
Then, in accordance with Sec.\ \ref{Reduction}, there are linear differential
relations between any $m\times r_k + 1$ of the functions 
$H(\vec{J}+\vec{m}; \vec{z})$. This means that these functions can be 
expressed in terms of a set of basis functions shifted by integer values of 
the parameters, 
\begin{equation}
H(\vec{J}+\vec{m};\vec{z}) = 
\sum_{K=0}^L R_K (\vec{z}) \frac{\partial^K}{\partial z_1^{k_1} \cdots \partial z_r^{k_r}} H_i(\vec{J};\vec{z}) \;.
\end{equation}

The number $L$ of elements on the r.h.s. of this relation is equal to the 
number of solutions of the corresponding differential equation.

\subsection{Application to Feynman Diagrams\label{Application}}
As an application of the reduction in Sec.\ \ref{Reduction}, a series of 
publications \cite{MKL06,BKKWY09,BKK2009,BKK2011} has analyzed the Feynman
diagrams $\Phi(n,\vec{j};z)$ having a
Mellin-Barnes representation of the form 
\begin{equation}
\Phi(n,\vec{j};z) 
= 
\frac{1}{2\pi i}
\int^{i\infty}_{-i\infty}
dt (\kappa z)^t \Gamma(-t) F(t)
\;, 
\label{MB}
\end{equation}
where $\vec{j}$ is the list of powers of the propagators in the Feynman diagram,
$n$ is the dimension of space-time, 
$\kappa$ is a a constant, 
and $F(t)$ has the structure
\begin{eqnarray}
F(t) = 
\prod_{i,j,k,l}
\frac{
\Gamma(A_i+t) 
\Gamma(C_k-t)
}
{
\Gamma(B_j+t) 
\Gamma(D_l-t)
}
\;, 
\label{F}
\end{eqnarray}
where $A_i, B_j, C_k, D_l$ are linear combinations of $n$ and $\vec{j}$ 
of the form
$
A_i = a_0 n + \sum_{k}a_k j_k  \;,
$
{\it etc.}, and the dimensions of the lists ${\vec A}, {\vec B}, {\vec C},
{\vec D}$ of these linear 
combinations satisfy the condition
\begin{equation}
\mbox{dim}\ \vec{A} + \mbox{dim}\ \vec{D} - \mbox{dim}\ \vec{B} - 
\mbox{dim}\ \vec{C} = 1 \;.
\label{dim}
\end{equation}
Closing the integration contour gives
\begin{equation}
\Phi(n,\vec{j};z) 
= \sum_{a=1}^q z^{l_a} C_{l_a}(n,\vec{j})
{\ }_{p+1}F_{p}(\vec{\widetilde{A}}_a;\vec{\widetilde{B}}_a; \kappa z) \;, 
\label{eq}
\end{equation}
where 
$q$ is an integer, $l_a, {\vec{\widetilde{A}}}_a, {\vec{\widetilde{B}}}_a$
are linear combinations of $n$ and $\vec{j}$ with rational coefficients, 
$z\ne 1$, and the coefficients $C_{l_a}$ are products of $\Gamma$-functions 
with arguments depending only on $n$ and $\vec{j}$. 

In accordance with the preliminary remarks in this section, 
the maximal number of basis elements 
for Eq.\ (\ref{eq}) is equal to $q \times p$. This number should coincide with 
the number of master integrals found by other means, in particular, by 
integration-by-parts (IBP) relations \cite{ibp}. However, some of the 
parameters of the hypergeometric functions on the r.h.s. of Eq.\ (\ref{eq}) may 
be exceptional. Typically, this occurs when the upper parameters are integers,
or the difference between an upper and lower parameter is a 
positive integer. \cite{BKK2009} In these cases, the actual number of 
basis elements will be less than the typical number.
(See also the discussion in Ref.\ \cite{MKL06}.)

The number of nontrivial basis elements for a hypergeometric function of one 
variable may be defined to be the highest power of the differential operator
$\theta = z d/dz$ in its differential reduction: \cite{BKK2009}
\begin{equation}
{}_{p+1}F_{p}(\vec{A};\vec{B};z) = 
\sum_s 
\left[ 
\sum_{l=0}^v P_l(z)\ \theta^{l} {\ }_{s+1}F_{s}
(\vec{A}-\vec{I}_1;\vec{B}-\vec{I}_2;z) + R_s(z)
\right] 
\;,
\label{eq2a}
\end{equation}
where $\vec{I}_1,\vec{I}_2$ are lists of integers and $P_l(z), R(z)$
are rational functions. On the other hand, applying the IBP relations to 
Eq.\ (\ref{eq}) would lead to an expression of the form
\begin{equation}
\Phi(n,\vec{j};z) = \sum_{k=1}^h B_k(n,\vec{j};z) \Phi_k(n;z) \;,  
\label{eq2b}
\end{equation}
where some of the master integrals $\Phi_k$ are expressible solely in terms of 
$\Gamma$-functions.\cite{vladimirov}

Analyzing the relation between the reductions (\ref{eq2a}) and (\ref{eq2b}) 
leads to the following criteria for determining the number of master 
integrals:\footnote{See Ref.\ \cite{BKK2009} for details. 
We thank Bas Tausk for bringing to our attention the fact that the diagram 
$F_2$ in that paper was analyzed in detail in Refs.\ 
\cite{Bonciani:1} (Eqs.~(74), (75)) and 
\cite{Bonciani:2} (Eqs.~(124), (130)). 
The results of these analyses are in agreement.}
\begin{itemize}
\item[(i)] Each term in the hypergeometric representation of a Feynman diagram
Eq.\ (\ref{eq}) has the same number $L$ of nontrivial basis elements (up to
rational functions). 
\item[(ii)] The number of master integrals following from the IBP relations 
which are not expressible solely in terms of 
$\Gamma$-functions \cite{vladimirov} is equal to the number $L$ of nontrivial 
basis elements.  
\end{itemize}

\subsection{Examples: Vertex-Type Diagrams\label{examples}} 
\noindent
In this section, we consider two examples of the differential reduction of 
one-loop vertex-type diagrams.

\par\noindent
{\bf Example I}: 
Let us consider the one-loop vertex
($C_3$ in the notation of Ref.\ \cite{BKK2009};
see Eq.~(52) there for details):
\begin{eqnarray} 
&& 
C^{(q)}_3(j_1,j_2,\vec{\sigma})
\equiv 
\int
\left.
\frac{d^nk}
{
[(k-p_1)^2-m^2]^{j_1}
[(k+p_2)^2-m^2]^{j_2}
(k^2)^{\vec{\sigma}}
}
\right|_{p_1^2=p_2^2=0}
\nonumber \\ &&
= i^{1-n} \pi^{n/2}
(-m^2)^{\tfrac{n}{2} - \sigma - j_{12}}
\frac{\Gamma\left( j_{12}  + \sigma -  \frac{n}{2} \right) 
      \Gamma\left( \frac{n}{2}  -  \sigma \right)}
     {\Gamma\left( j_{12} \right)
      \Gamma\left( \frac{n}{2} \right) }
{\ }_{4}F_3\left(\begin{array}{c|}
j_{12}  + \sigma -  \tfrac{n}{2}, j_1, j_2, \tfrac{n}{2}  -  \sigma \\
\tfrac{n}{2},  \tfrac{j_{12}}{2}, \tfrac{j_{12} + 1}{2} \end{array} 
~\frac{(p_1-p_2)^2}{4m^2} \right) 
\;, 
\label{C3}
\end{eqnarray} 
where $j_{12} \equiv j_1 + j_2$, $\sigma \equiv \sum_{k=1}^{q+1} \sigma_k, $
and the ``dressed'' massless propagator is 
\begin{eqnarray}
\frac{1}{(k^2)^{\vec{\sigma}}}
\equiv 
\left\{ \prod_{k=1}^{q+1} \frac{\Gamma(\frac{n}{2}-\sigma_k)}{\Gamma(\sigma_k)} \right\}
\frac{
\left[ i^{1-n} \pi^{n/2} \right]^q
\Gamma\left(\sigma-\frac{n}{2}q\right)
}{
\Gamma\left(\frac{n}{2}(q+1) - \sigma \right)} 
\frac{1}{(k^2)^{\sigma-\tfrac{n}{2}q}}
\; ,
\label{sigma}
\end{eqnarray}
where $q$ is the number of massless loops, $(q+1)$ is the number of 
massless lines, and  $q=0$ corresponds to a massless line without 
additional internal loops. ($C_3^{(0)}$ is a true one-loop vertex.)

With the redefinition $\sigma \to \sigma - \tfrac{n}{2}q$, 
the hypergeometric function in Eq.~(\ref{C3}) takes the form
\begin{eqnarray}
{}_{4}F_3\left(\begin{array}{c|}
j+\sigma  -  \tfrac{n}{2}(q+1), j_1, j_2, \tfrac{n}{2}(q+1)  -  \sigma \\
\tfrac{n}{2},  \tfrac{j}{2}, \tfrac{j + 1}{2} \end{array} ~\frac{(p_1-p_2)^2}{4m^2} \right) 
\;.
\end{eqnarray}
In accordance with the differential reduction algorithm, 
this function may be expressed in terms of a ${}_{3}F_2$ function with 
one unit upper parameter,
and its first derivative,
\begin{equation}
\{ 
1, \theta
\}
\times
{}_{3}F_2\left(\begin{array}{c|}
1, \tfrac{n}{2}(q+1) + I_1, I_2 - \tfrac{n}{2}(q+1) \\
\tfrac{n}{2}, \tfrac{1}{2} + I_3  \end{array} ~z \right) \;, 
\label{C3:2}
\end{equation}
together with rational functions of $z$, for integer values $I_1, I_2, I_3$.
The short-hand notation $(1, \theta)$ stands for a 
combination $P_1(z) + P_2(z)\theta$, with rational functions $P_i$.
(See Eqs.~(17) and (20) in Ref.\ \cite{BKK2009}.)
According to the criteria in Sec.\ \ref{Application}, there are two master 
integrals for this diagram which are not expressible in terms of $\Gamma$ 
functions.  For $q= 1$, the standard approach based on  IBP
relations, \cite{ibp} yields two two-loop vertex master integrals.
(See Eqs.~(3) and (9) in Ref.\ \cite{2vertex}.) 
These master integrals are relevant for the massless fermion contribution 
to Higgs production and decay. \cite{Hgg}
It is interesting to note that at the one-loop level, $C^{(0)}_3$ has one 
master integral of the vertex type and one of the propagator type, 
which are again equivalent to Eq.~(\ref{C3:2}). 
\\[1ex]

\par\noindent
{\bf Example II}: 
Let us consider one-loop vertex diagram $C_1^{(q_1,q_2)}$, defined as 
\begin{eqnarray} 
&& 
C^{(q_1,q_2)}_1(\vec{\sigma}_1,\vec{\sigma}_2,\rho)
\equiv 
\int
\left.
\frac{d^nk}
{
[(k-p_1)^2]^{\vec{\sigma}_1}
[(k+p_2)^2]^{\vec{\sigma}_2}
(k^2-m^2)^{\rho}
}
\right|_{p_1^2=p_2^2=0}
\;, 
\label{C1}
\end{eqnarray} 
where 
$
\vec{\sigma}_1, \vec{\sigma}_2
$
are defined as in Eq.~(\ref{sigma}), with 
$\sigma_j = \sum_{k=1}^{q_j+1} \sigma_{jk}$ for $j=1,2$. 
The case $q_1=0,q_2=1$ corresponds to Eq.~(173) in Ref.\ \cite{Bonciani:3}. 
The hypergeometric representation for this diagram is \cite{AGO2}
\begin{eqnarray} 
&& 
C^{(q_1,q_2,)}_1(\vec{\sigma}_1,\vec{\sigma}_2,\rho)
= 
i^{1-n} \pi^{n/2}
(-m^2)^{\tfrac{n}{2} - \rho - {\sigma_1} - {\sigma_2}}
\nonumber \\ && \hspace{5mm}
\Biggl\{
\frac{\Gamma\left( \rho \!+\! {\sigma_1} \!+\! {\sigma_2}  
	\!-\!  \frac{n}{2} \right) 
      \Gamma\left( \frac{n}{2}  \!-\!  {\sigma_1} 
	\!-\! {\sigma_2} \right)}
     { \Gamma\left( \frac{n}{2} \right) \Gamma(\rho)}
{\ }_{3}F_2\left(\begin{array}{c|}
 \rho \!+\! {\sigma_1} \!+\! {\sigma_2}  
	\!-\!  \tfrac{n}{2}, {\sigma_1}, {\sigma_2} \\
\tfrac{n}{2}, 1  \!+\!  {\sigma_1} \!+\! {\sigma_2}   
	\!-\!  \tfrac{n}{2}  
\end{array} ~-\frac{(p_1-p_2)^2}{m^2} \right) 
\nonumber \\ && \hspace{5mm}
+ \left( - \frac{(p_1-p_2)^2}{m^2}\right)^{\tfrac{n}{2} 
\!-\! {\sigma_1} \!-\! {\sigma_2} }
\frac{\Gamma\left( \frac{n}{2}  \!-\!  {\sigma_1} \right) 
      \Gamma\left( \frac{n}{2}  \!-\!  {\sigma_2} \right) 
      \Gamma\left( {\sigma_1} \!+\!  {\sigma_2}    
	\!-\!  \frac{n}{2} \right)}
     { \Gamma\left( n  \!-\! {\sigma_1} \!-\!  {\sigma_2}  \right) 
	\Gamma({\sigma_1}) \Gamma({\sigma_2} )}
\\ && \hspace{10mm}
\times
{}_{3}F_2\left(\begin{array}{c|}
\rho, \tfrac{n}{2} \!-\! {\sigma_1}, \tfrac{n}{2} \!-\! {\sigma_2}  \\
n \!-\! {\sigma_1} \!-\! {\sigma_2} , 
\tfrac{n}{2}  \!-\!  {\sigma_1}  \!-\! {\sigma_2}   \!+\!  1 \end{array}
 ~ -\frac{(p_1-p_2)^2}{m^2} \right) 
\Biggr\} \;.\nonumber
\label{C1:2}
\end{eqnarray}
In accordance with the differential reduction algorithm, 
the first ${}_3F_2$ function in Eq.~(\ref{C1:2}) can be expressed in
terms of ${}_{2}F_1$, and the second one can be expressed in terms of 
${}_3F_2$ with one unit upper parameter, namely 
\begin{eqnarray}
&&
\{ 
1, \theta
\}
\times
{}_{2}F_1\left(\begin{array}{c|}
{\sigma_1}, {\sigma_2} \\
\tfrac{n}{2}, 
\end{array} ~z \right) 
\;, 
\quad 
\{ 
1, \theta
\}
\times
{}_{3}F_2\left(\begin{array}{c|}
1, \tfrac{n}{2} \!-\! {\sigma_1}, \tfrac{n}{2} \!-\! {\sigma_2}  \\
n \!-\! {\sigma_1} \!-\! {\sigma_2} , 
\tfrac{n}{2}  \!-\!  {\sigma_1}  \!-\! {\sigma_2}   \!+\!  1 
\end{array} ~z \right)  \;,
\label{C1:3}
\end{eqnarray}
together with rational functions of $z$. We note that each hypergeometric 
function in Eq.\ \ref{C1:2} has the same number of nontrivial basis elements, 
in agreement with criterion (i) in section \ref{Application}.
When one (or both) of the ${\sigma_j}$ are integers ($q_j=0$), further 
reduction is possible to
\begin{eqnarray}
&&
{}_{2}F_1\left(\begin{array}{c|}
1, {\sigma_2} \\
\tfrac{n}{2}, 
\end{array} ~z \right) 
\;, 
\quad 
{}_{2}F_1\left(\begin{array}{c|}
1,  \tfrac{n}{2} \!-\! {\sigma_2}  \\
n \!-\! {\sigma_1} \!-\! {\sigma_2}  
\end{array} ~z \right)  \;.
\label{C1:4}
\end{eqnarray}
In accordance with our criteria, there are two nontrivial master integrals for 
non-integer $\sigma_j$, and one nontrivial master integral when one of the 
$\sigma_j$ is an integer. The last result in agreement with 
Ref.\ \cite{Bonciani:3}.

\newpage
\subsection{Counting the Non-Trivial Master Integrals\label{counting}}
When analyzing various diagrams, one example was found \cite{KK2011} where 
our criteria allow us to predict and prove an additional relation between 
master integrals.  It is a two-loop sunset diagram relevant for the 
evaluation of ${\cal O}(\alpha \alpha_s)$ relation between the pole
and $\overline{\rm MS}$ mass of the top-quark in the Standard 
Model, \cite{JK2003} as well as in the gaugeless-limit model. \cite{JK2004}
To illustrate this, let us consider the two-loop self-energy diagram  
\begin{eqnarray}
&& 
V_{1200}(\rho,\sigma, \alpha, \beta, m^2,M^2) 
= 
\left. 
\int 
\frac{d^n (k_1 k_2)}{
[(k_1\!-\!p)^2\!-\!m^2]^\rho [(k_1\!-\!k_2)^2-M^2]^\sigma 
	[k_2^2]^\alpha [k_1^2]^{\beta} 
}
\right |_{p^2 = m^2} 
\;,
\label{V1200}
\end{eqnarray}
which is generated in an intermediate step 
of the calculations in Ref.\ \cite{JK2004}. 
The $\alpha=0$ case corresponds to a diagram considered in Ref.\ \cite{KK2011}.
The Mellin-Barnes representation of the integral (\ref{V1200}) is 
\begin{eqnarray}
&& 
V_{1200}(\rho,\sigma, \alpha, \beta, m^2,M^2) 
= 
[i^{1-n} \pi^\frac{n}{2}]^2
\frac{\Gamma(\frac{n}{2}-\alpha)}{\Gamma(\alpha) \Gamma(\rho) \Gamma(\sigma)}
(-m^2)^{n-\alpha-\beta-\sigma-\rho}
\int ds  \left( \frac{M^2}{m^2} \right)^s 
\nonumber \\ && 
\frac{
\Gamma(-s) 
\Gamma\left(\frac{n}{2} \!-\! \sigma  \!-\! s \right) 
\Gamma\left(2n \!-\! 2 \alpha \!-\! 2 \beta \!-\! 2 \sigma  \!-\! \rho \!-\! 2s \right) 
\Gamma\left(\alpha \!+\! \beta  \!+\! \sigma \!+\! \rho \!-\! n \!+\! s \right) 
\Gamma\left(\alpha \!+\! \sigma \!-\! \frac{n}{2} \!+\! s \right) 
}
{
\Gamma\left(n \!-\! \alpha  \!-\! \sigma \!-\! s \right) 
\Gamma\left(\frac{3n}{2} \!-\! \alpha  \!-\! \beta \!-\! \sigma \!-\! \rho\!-\! s \right) 
}
\;.
\end{eqnarray}
After integration, we obtain
\begin{eqnarray}
&& 
V_{1200}(\rho,\sigma, \alpha, \beta, m^2,M^2) 
= 
[i^{1-n} \pi^\frac{n}{2}]^2
\frac{\Gamma(\frac{n}{2}-\alpha)}{\Gamma(\alpha) \Gamma(\sigma) 
	\Gamma\left(\frac{n}{2}\right)}
(-M^2)^{n-\alpha-\beta-\sigma-\rho}
\nonumber \\ & 
\times&
\Biggl[ 
\left( 
\frac{m^2}{M^2}
\right)^{\frac{n}{2} - \beta - \rho}
\frac{
\Gamma(\alpha)
\Gamma\left(\alpha \!+\! \sigma \!-\! \frac{n}{2}\right)
\Gamma\left(\beta \!+\! \rho \!-\! \frac{n}{2}\right)
\Gamma(n \!-\! 2\beta \!-\! \rho)
}{
\Gamma(\rho)
\Gamma(n\!-\!\beta\!-\!\rho)
}
~{}_{4}F_3 \left(\begin{array}{c|}
\alpha, \alpha \!+\! \sigma \!-\! \frac{n}{2}, \frac{n-\rho}{2} \!-\! \beta,  
	\frac{n+1-\rho}{2} \!-\! \beta \\
\frac{n}{2}, n\!-\!\beta\!-\!\rho, \frac{n}{2}+1\!-\!\beta\!-\!\rho 
\end{array} ~\frac{4 m^2}{M^2} \right)
\nonumber \\ &+&
\frac{
\Gamma\left(\alpha \!+\! \beta \!+\! \sigma \!+\! \rho \!-\! n \right)
\Gamma\left(\alpha \!+\! \beta \!+\! \rho \!-\! \frac{n}{2}\right)
\Gamma\left(\frac{n}{2} \!-\! \beta \!-\! \rho\right)
}{
\Gamma(\beta\!+\!\rho)
}
~{}_{4}F_3 \left(\begin{array}{c|}
\alpha \!+\! \beta \!+\! \sigma \!+\! \rho \!-\! n, 
\alpha \!+\! \beta \!+\! \rho \!-\! \frac{n}{2}, 
\frac{\rho}{2},  \frac{\rho+1}{2} \\
\frac{n}{2}, 
\beta\!+\!\rho, 
1\!+\!\beta\!+\!\rho \!-\! \frac{n}{2} 
\end{array} ~\frac{4 m^2}{M^2} \right)
\Biggr].
\label{V1200:hyper}
\end{eqnarray}
The differential reduction of the hypergeometric functions 
in Eq.~(\ref{V1200:hyper}) can be expressed as 
\begin{eqnarray}
&& 
~{}_{4}F_3 \left(\begin{array}{c|}
\alpha, \alpha \!+\! \sigma \!-\! \frac{n}{2}, \frac{n-\rho}{2} \!-\! \beta,  
	\frac{n+1-\rho}{2} \!-\! \beta \\
\frac{n}{2}, n\!-\!\beta\!-\!\rho, \frac{n}{2}+1\!-\!\beta\!-\!\rho 
\end{array} ~z\right)
\to 
(1, \theta)
\times
{}_{3}F_2 \left(\begin{array}{c|}
1,
I_1 \!-\! \tfrac{n}{2},
\tfrac{n}{2} \!+\! \tfrac{1}{2} \!+\! I_2 \\
n \!+\! I_3, 
\tfrac{n}{2} \!+\! I_4
\end{array} ~z \right) \!+\! R_1(z) \;,
\nonumber\\ && 
~{}_{4}F_3 \left(\begin{array}{c|}
\alpha \!+\! \beta \!+\! \sigma \!+\! \rho \!-\! n, 
\alpha \!+\! \beta \!+\! \rho \!-\! \frac{n}{2}, 
\frac{\rho}{2},  \frac{\rho+1}{2} \\
\frac{n}{2}, 
\beta\!+\!\rho, 
1\!+\!\beta\!+\!\rho \!-\! \frac{n}{2} 
\end{array} ~z \right)
\to 
(1, \theta)
\times
{}_{3}F_2 \left(\begin{array}{c|}
I_1 \!-\! n,
\tfrac{1}{2} \!+\! I_2,1 \\
\tfrac{n}{2} \!+\! I_3,2 
\end{array} ~z \right)
\;.
\end{eqnarray}
In accordance with our criteria, 
there should exist two algebraically independent master integrals 
and some integrals expressible in terms of rational functions. 

Diagram (\ref{V1200}) is algebraically reducible via IBP relations 
\cite{Tarasov} to four new diagrams with 
(i) $\sigma=0$,
(ii) $\rho=0$,
(iii) $\alpha=0$,
(iv) $\beta=0$.
The $\sigma=0$ diagram vanishes in framework of dimensional regularization. 
The diagrams with $\beta=0$ and $\alpha=0$ are irreducible, and treated as two 
independent master integrals. 
The diagram with $\beta=0$ is a two-loop sunset, and in accordance with 
the IBP algorithm of Ref.\ \cite{Tarasov}, it has 
three master integrals of the same topology plus a product of two one-loop 
bubbles $(\alpha=\beta=0)$. 
In this way, the classical IBP relations applied to diagram (\ref{V1200})
give six master integrals, without any information about its 
algebraic structure.  The hypergeometric representation for sunset diagrams 
(\ref{V1200:hyper}) and the differential reduction algorithm 
(\ref{decomposition:1b}) allow us to find algebraic relations between 
the three sunset-type master integrals. (See Eq.(9) in Ref.\ \cite{KK2011}.)
The on-shell case of a new relation (Eq.(10) in Ref.\ \cite{KK2011}) enters 
in the package {ON-SHELL2} \cite{onshell2},
and was postulated via a study of the higher-order coefficients 
of the $\ep$-expansion in Ref.\ \cite{FK99}.  In Ref.\ \cite{czakon}, it 
was pointed out that the last on-shell relation can 
be treated purely diagrammatically.  In this case, the new diagram does not 
follow from the original sunset by the contraction of any lines.  Surprisingly, 
this new relation is not reproduced by the IBP algorithm of 
Ref.\ \cite{Tarasov} or by the automated programs {AIR} \cite{AIR} and 
{FIRE} \cite{FIRE}.  This may be an artifact of 
the ``topological'' nature of the solution of IBP relations (systematic 
contractions of one line in a diagram). 
The effect that the solution of IBP relations does not recognize the 
algebraic relation between master integrals
is also seen for phase-space integrals: some of the master integrals 
collected in Ref.\ \cite{primitive} are also algebraically related to 
each other. 


\section{Construction of the $\ep$ Expansion\label{Expansion}}
The evaluation of multi-loop radiative corrections requires higher-order
terms in the $\ep$ expansion of lower-order diagrams.
The integral representation \cite{AGO2}, 
the series representation \cite{MUW,JKV,DK04,Weinzierl:2004,series:all-order}
or the differential equations approach \cite{KWY,KK1,KK2,KK3} may be used to 
construct the $\ep$ expansion of hypergeometric functions. We will focus
on the differential equations approach. As is well-known,
hypergeometric functions satisfy a differential equation
\begin{eqnarray}
\left[
z \Pi_{i=1}^{p }\left( z \frac{d}{dz} \!+\! A_i \right) 
\!-\!  z \frac{d}{dz} \Pi_{k=1}^{p-1} \left( z \frac{d}{dz} 
	\!+\! B_k\!-\!1 \right) 
\right] {}_{p}F_{p-1}(\vec{A};\vec{B}; z) = 0.
\label{diff:eq}
\end{eqnarray}
Due to the analyticity of the hypergeometric function 
${}_{p}F_{p-1}(\vec{A};\vec{B}; z)$ 
with respect to its parameters $A_i,B_k$, \
the differential equation for the coefficient functions $\omega_k$
of the  Laurent expansion can be derived directly from Eq.~(\ref{diff:eq}) 
without any reference to the series or integral representation.  
This was the main idea of the approach developed in 
Refs.\ \cite{KWY,KK1,KK2}.

It is convenient to introduce a new parametrization, 
$
A_i \to A_{i}+a_i\ep, B_j \to B_{i}+b_i\ep \;,
$
where $\ep$ is small, so that the Laurent expansion 
takes the form of an ``$\ep$ expansion,'' 
$$
\omega(z) \equiv 
{}_{p}F_{p-1}(\vec{A}+\vec{a}\ep;\vec{B}+\vec{b}\ep; z)
 = 
{}_{p}F_{p-1}(\vec{A};\vec{B}; z)
+ \sum_{k=1}^\infty \ep^k \omega_k(z) 
\;.
$$
We can rewrite Eq.~(\ref{diff:eq}) as a system of first-order 
differential equations (called the ``Pfaff form'') and expand all terms
in powers of $\ep$: 
\begin{eqnarray}
d \phi^{(i)}(z,\ep) = \sum_{j=0}^{p-1} P_{i,j}(z,\ep) \phi^{(j)}(z,\ep)dz \;,
\label{diff:pfaff}
\end{eqnarray}
where $\phi^{(i)}(z) = h_i(z) \theta \phi^{(i-1)}(z), i = 1, \cdots, p-1$ and 
$\phi^{(0)} = \omega(z)$, with arbitrary functions $h_i(z)$.
The result can be written in the form
\begin{eqnarray}
P_k(z) \frac{d}{dz} f^{(k)}_j(z) = \sum_{m,l} R^{(k,m)}_{j,l}(z) 
f^{(m)}_{j-1-l}(z) \;, \quad j = 0, \cdots, \infty\;,
\label{expanded}
\end{eqnarray}
where $P_k(z)$ and $R^{(k,m)}_{j,l}(z)$ are polynomials. 
For a restricted set of parameters, the expanded system of differential 
equations in Eq.~(\ref{expanded}) has triangular form and can be integrated
iteratively in terms of Goncharov's polylogarithms.\footnote{The fact 
that higher-order differential operators depending on $\ep$ may take 
a simpler form after $\ep$-expansion has also been observed in the evaluation 
of multi-loop master integrals; see {\it e.g.} 
Refs.\ \cite{sunset,Bonciani:1,Bonciani:2,Bonciani:3,Bonciani:4,2vertex}}

For illustration, let us consider the factorization of a
differential operator expanded in powers of $\ep$:
\begin{eqnarray}
D^{(p)} = 
\left[
\Pi_{i=1}^{p }\left( \theta \!+\! A_i \!+\! a_i \ep\right) 
\!-\!  \frac{1}{z} \theta \Pi_{k=1}^{p-1} \left( \theta \!+\! B_k\!-\!1 
	\!+\! b_k\ep\right) 
\right] = 
\sum_{j=0}^{p} \ep^j D_j^{(p-j)}(\vec{A},\vec{B},\vec{a},\vec{b},z)  \;, 
\end{eqnarray}
where $\theta = z d/dz$,
the upper index gives the order of the differential operator, 
$
D_p^{(0)} = \Pi_{k=1}^p a_k \;,
$
and 
\begin{eqnarray}
D_0^{(p)} & = &   
\Pi_{i=1}^{p }\left( \theta \!+\! A_i \right) 
\!-\!  \frac{1}{z} \theta \Pi_{k=1}^{p-1} \left( \theta \!+\! B_k\!-\!1 \right) 
\\ 
 & = &  
\left\{ -(1\!-\!z) \frac{d}{dz}   
\!+\! \sum_{k=1}^p A_k \!-\! \frac{1}{z} \sum_{j=1}^{p-1} (B_j\!-\!1)
\right\} \theta^{p-1} 
\!+\! 
\sum_{j=1}^{p-2} 
\left[ P^{(p)}_{p-j}(\vec{A}) 
\!-\! \frac{1}{z} P^{(p-1)}_{p-j}(\vec{B}-\vec{1}) 
\right] \theta^j
+ \sum_{i=0}^p A_i
\;, 
\nonumber  
\label{D0}
\end{eqnarray}
with polynomials $P^{(p)}_j(r_1,\cdots,r_p)$ defined 
via the relations
\begin{equation}
\prod_{k=1}^{p}(z+r_k) = 
\sum_{j=0}^{p} P^{(p)}_{p-j}(r_1,\cdots,r_p) z^j 
\equiv \sum_{j=0}^p P^{(p)}_{p-j}(\vec{r}) z^j 
\equiv \sum_{j=0}^p P^{(p)}_{j}(\vec{r}) z^{p-j} \;,
\label{P}
\end{equation}
and $\vec{1}_k \equiv 1$.
As a first step toward the construction of a solution, the operator 
$D_0^{(p)}$ should be rewritten in the form
\begin{eqnarray}
D_0^{(p)}  = 
\left\{ -(1\!-\!z) \frac{d}{dz}   
\!+\! R_1 \!-\! \frac{1}{z} R_2
\right\} 
\Pi_{j=1}^{p-1} \left( \theta + \beta_j  \right) \;,
\label{factorization}
\end{eqnarray}
where $\beta_j$ are rational numbers. 
Eqs.~(\ref{P}) and (\ref{factorization}) give rise to a system of equations 
\begin{eqnarray}
&& 
R_2 P^{p-1}_{k-1}(\vec{\beta}) + P^{p-1}_{k}(\vec{\beta}) 
= P^{p-1}_k (\vec{B}-\vec{1}) \;, \quad k = 1, \cdots, p \;.\nonumber
\\ && 
R_1 P^{p-1}_{k-1}(\vec{\beta}) + P^{p-1}_{k}(\vec{\beta}) 
= P^{p}_k (\vec{A}) \;, \quad k = 1, \cdots, p \;.
\label{R2-R1}
\end{eqnarray}
The differential equation 
\begin{equation}
\left\{ -(1\!-\!z) \frac{d}{dz}   \!+\! R_1 \!-\! \frac{1}{z} R_2 \right\} h(z)
 = 0 \;,
\label{diff-eq}
\end{equation}
generates the function 
\begin{equation}
h(z) = C z^{-R_2} (z-1)^{R_2-R_1} \;,
\label{h}
\end{equation}
for which only three rational parametrizations are known:
\begin{equation}
\hbox{(i)} R_1 = R_2; 
\qquad 
\hbox{(ii)}
R_1 = 0; \qquad 
\hbox{(iii)}
R_2 = 0 \;.
\label{R-relations}
\end{equation}

When all $A_i=0$ and  $B_j=1$, we have integer parameters and the 
differential operators are factorizable. 
When only one upper parameter $A_1 \neq 0$ and one lower parameter 
$B_1 \neq 1$, the $\beta_j$ again all vanish, and  
a parametrization $z \to \xi(z)$ should exist such that 
$
[(1-z) h(z)]^{-1} dz/d\xi, 
\ [z h(z)]^{-1} dz/d\xi, 
\ z^{-1} dz/d\xi, 
$
are rational functions of $\xi$, 
where $h(z) = C z^{1-B_1}(z-1)^{B_1-A_1-1}$. (See Ref.\ \cite{KK2}.)\\[1ex]

\par\noindent 
{\bf Example III:} 
Let us consider 
a Gauss hypergeometric function of the form
$
\omega(z) = 
~{}_2F_1(
\frac{p_1}{q}+a_1 \ep,
\frac{p_2}{q}+a_2 \ep;$ $
1-\frac{r}{q}+c \ep; z 
) \;,
$
where $p_1,p_2,q,r$ are integers.
It is a solution of the differential equation 
\begin{eqnarray}
\left( z \frac{d}{dz} + \frac{p_1}{q} + a_1 \ep \right) 
\left( z \frac{d}{dz} + \frac{p_2}{q} + a_2 \ep \right) 
\omega(z)
= 
\frac{d}{dz} 
\left( z \frac{d}{dz} - \frac{r}{q} + c \ep \right) \omega(z) 
\label{gauss:diff}
\end{eqnarray}
for coefficient functions $\omega_k(z)$ defined via the expansion
$
\omega(z) = 1+\sum_{k=1}^\infty \omega_k(z) \ep^k.
$
Eq.~(\ref{gauss:diff}) produces an infinite system of linear 
differential equations
\begin{eqnarray}
&& 
\hspace{-2em}\left[ 
(1-z) \frac{d}{dz} 
+ \left( \beta - \frac{p_1+p_2}{q} \right)
- \frac{1}{z} \left( \beta \!+\! \frac{r}{q} \right) \right] 
 \left( z \frac{d}{dz} \!+\! \beta \right) \omega_k
-
\left[ 
\left( 
\beta \!-\! \frac{p_1}{q}
\right)
 \left( 
\beta-\frac{p_2}{q}
\right) 
\!-\! \frac{1}{z} \beta \left( \beta \!+\! \frac{r}{q} \right) 
\right]
\omega_k(z)
\nonumber \\&&\hspace{-1em}=
\left(  a_1 \!+\! a_2  \!-\! \frac{c}{z} \right) 
 \left( z \frac{d}{dz} \!+\! \beta \right)
\omega_{k-1}(z)
+ \frac{c}{z} \beta \omega_{k-1}
- \left[ 
a_1 \left( \beta - \frac{p_2}{q} \right) 
+ a_2 \left( \beta - \frac{p_1}{q} \right) 
\right]\omega_{k-1}(z)
\!+\! 
a_1 a_2 \omega_{k-2}(z),\quad\strut
\label{gauss:factor}
\end{eqnarray}
where $\beta$ is arbitrary and $k$ runs from $0$ to $\infty$.
The second order differential equation (\ref{gauss:factor}) can be split
into two first-order differential equations by introducing 
new functions 
\begin{equation}
\rho_k(z)  =  \left( z \frac{d}{dz} \!+\! \beta  \right)  \omega_{k}(z) \;, 
\label{rho}
\end{equation}
so that 
\begin{eqnarray}
&& \left[ 
(1-z) \frac{d}{dz} 
+ \left( \beta - \frac{p_1+p_2}{q} \right)
- \frac{1}{z} \left( \beta + \frac{r}{q} \right) \right] 
\rho_k(z)
-
\left[ 
\left( 
\beta-\frac{p_1}{q}
\right)
 \left( 
\beta-\frac{p_2}{q}
\right) 
- \frac{1}{z} \beta \left( \beta + \frac{r}{q} \right) 
\right]
\omega_k(z)
\nonumber \\ &&\quad = 
\left(  a_1 \!+\! a_2  \!-\! \frac{c}{z} \right) 
\rho_{k-1}(z)
+ \frac{c}{z} \beta \omega_{k-1}
- a_1 \left( \beta - \frac{p_2}{q} \right) \omega_{k-1}(z)
- a_2 \left( \beta - \frac{p_1}{q} \right) \omega_{k-1}(z)
\!+\! 
a_1 a_2 \omega_{k-2}(z).\quad 
\label{gauss-rho-omega}
\end{eqnarray}
We can use the fact that this  system takes 
triangular form when the last term in the l.h.s.\ of the 
second equation is zero to obtain the following solutions:
\begin{eqnarray}
&& 
p_1 p_2  =  0 \longrightarrow \beta = 0 \;, 
\label{p1p2=0}
\\ &&
p_1  =  0 \longrightarrow \beta = 0 \;, 
\label{p1=0}
\\ && 
\beta = - \frac{r}{q} = \frac{p_1}{q} \;, \quad p_2 = p \;.
\label{p2}
\end{eqnarray}

We would like to analyze the case of the parameters in Eq.~(\ref{p2}) in 
detail. Then
\begin{eqnarray}
&& 
\rho_k(z)  =  \left( z \frac{d}{dz} \!-\! \frac{r}{q}  \right) \omega_{k}(z)\;, 
\\ && 
\left[ (1-z) \frac{d}{dz} \!-\! \frac{p}{q} \right] \rho_k(z)
= \left(  a_1 \!+\! a_2  \!-\! \frac{c}{z} \right) 
\rho_{k-1}(z)
\!+\! \frac{c}{z} \frac{p_1}{q} \omega_{k-1}(z)
\!-\! a_1 \frac{p_1-p}{q} \omega_{k-1}(z)
\!+\!  a_1 a_2 \omega_{k-2}(z) \;.
\label{gauss-rho-omega:1}
\end{eqnarray}
The redefinition $(\omega_k, \rho_k) \to \left( z^{\frac{r}{q}} \pi_k, 
	(1-z)^{-\frac{p}{q}} \sigma_k \right) $
leads to a new set of equations 
\begin{eqnarray}
&& 
\sigma_k(z)  =   h(z) z\frac{d}{dz}\pi_{k}(z) \;, 
\\ && 
(1\!-\!z) \frac{d}{dz}  \sigma_k(z)
= 
\left(  a_1 \!+\! a_2  \!-\! \frac{c}{z} \right) 
\sigma_{k-1}(z)
+ \frac{c}{z} \frac{p_1}{q} h(z) \pi_{k-1}(z)
\!-\! a_1 \frac{p_1-p}{q} h(z) \pi_{k-1}(z)
\!+\! 
a_1 a_2 h(z) \pi_{k-2}(z),
\label{gauss-rho-omega:2}
\end{eqnarray}
where $h(z) =  (1-z)^{p/q} z^{r/q}$.
The result can be expressed in terms of Goncharov's 
polylogarithms if there is a parametrization $z \to \xi(z)$ 
such that $[z h(z)]^{-1} dz/d\xi$, $[(1-z) h(z)]^{-1} dz/d\xi$, 
$z^{-1} dz/d\xi$, and $(1-z)^{-1} dz/d\xi$
are rational functions. (See Ref.\ \cite{KK2}.)
Such a parametrization exists when $p=-r$ and 
$z \to \xi = \left(\frac{z}{z-1} \right)^{1/q}.$
In this way, we find the following set of conditions:
\begin{equation}
\beta = - \frac{r}{q} = \frac{p_1}{q} = \frac{p_2}{q} \;. 
\label{p2:2}
\end{equation}
Under these conditions, the coefficients of the $\ep$-expansion of 
the function 
$
~{}_2F_1(
\frac{p_1}{q}+a_1 \ep,
\frac{p_2}{q}+a_2 \ep;
1-\frac{r}{q}+c \ep; z 
) \;,
$
are expressible in term of Goncharov's polylogarithms. 
Eq.~(\ref{p2:2}) corresponds to Lemma IV of Ref.\ \cite{KK1}, but the present
derivation does not rely on any 
symmetries of Gauss hypergeometric functions. \\[1ex]

\par\noindent
{\bf Remark 1:} For a Gauss hypergeometric function, the $\ep=0$ term 
should be a rational (not just algebraic) function. \\[1ex]

\par\noindent
{\bf Remark 2:} 
For Eq.~(\ref{p1=0}), an additional relation between $p_2$ and $r$ arises from
Eq.~(\ref{diff-eq}), and we arrive again at
one of the cases of Eq.~(\ref{R-relations}).\\[1ex]

Let us consider the $\ep$ expansion of a hypergeometric function with
the following set of parameters:
$
\omega(z) = 
{}_pF_{p-1}\left( \vec{a} \ep, A_1\!+\!c_1 \ep, A_2\!+\!c_2 \ep; 
        \vec{b} \ep, B_1\!+\!\!f_1 \ep, B_2\!+\!\!f_2 \ep;z\right)
$,
where $A_1,A_2$, $B_1,B_2$, $\vec{a}$, $\vec{b}$,
$c$, and $f$ are rational numbers. 
Eqs.\ (\ref{R2-R1}) take the form 
\begin{eqnarray}
R_2 + \beta & = & (B_1-1) + (B_2-1)\;, 
\qquad 
R_1 + \beta = A_1 + A_2\;, 
\nonumber \\ 
R_2 \beta & = & (B_1-1) (B_2-1) \;, 
\qquad \qquad
R_1 \beta = A_1 A_2 \;. 
\end{eqnarray}
Eq.\ (\ref{R-relations}) provides additional conditions on the 
relations between $R_1$ and $R_2$, and as a
consequence, there are three different solutions:
\begin{eqnarray} 
R_1 = R_2 \;: \quad & & 
B_1-1 + B_2 -1 = A_1 + A_2 \;, 
\quad 
(B_1-1) (B_2-1) = A_1 A_2;
\label{I}\\[1ex]
R_1 = 0\;: \quad & & 
A_2 = 0\;, 
\quad 
\beta = A_1 \; ,
\nonumber \\ &&
R_2 = (B_1-1) + (B_2-1) - A_1,
\quad 
R_2 A_1  = (B_1-1)(B_2-1);
\label{II}\\[1ex]
R_2 = 0\;:  \quad
& & 
B_2 = 1\;, 
\quad 
\beta = B_1 \!-\! 1 \; 
\nonumber \\ &&
R_1 = A_1 + A_2 - (B_1-1) \;,
\quad 
R_1 (B_1-1)  = A_1 A_2;
\label{III}
\end{eqnarray}
The solutions of Eqs.~(\ref{I})  and (\ref{III}) 
are the roots of a quadratic equation $x^2-(A_1+A_2)x+A_1A_2=0$,
and the solution of Eq.~(\ref{II}) satisfies the same 
quadratic equation with the replacement $A_i \to B_i-1$.
One solution of Eq.~(\ref{I}) is $B_j = 1+A_j$,
and one solution of Eqs.~(\ref{II}) and (\ref{III}) is $B_1= 1+A_1$.

There is another parametrization
for the same hypergeometric function.
Let us rewrite the operator $D_0^{(p)}$ in Eq.~(\ref{D0}) as 
\begin{eqnarray}
&& 
\left\{ 
\left[ 
-(1\!-\!z) \frac{d}{dz}   
\!+\! \sum_{k=1}^2 A_k \!-\! \frac{1}{z} \sum_{j=1}^{2} (B_j\!-\!1)
\right] \theta 
+ A_1 A_2 - \frac{1}{z} (B_1-1) (B_2-1) 
\right\}
\theta^{\,p-2} 
\nonumber \\ && \qquad
= 
\left\{ 
\left[ 
-(1\!-\!z) \frac{d}{dz}   
\!-\! \left( \beta \!-\! A_1 \!-\! A_2 \right) \!+\! \frac{1}{z} 
	\left(\beta \!-\! B_1 \!-\! B_2 \!+\! 2 \right)
	\right] \left( \theta \!+\! \beta \right) \right\} \theta^{\,p-2} 
\label{D0:2} 
\\ &&  \qquad
+ 
\left\{ 
\left( \beta - A_1 \right) \left( \beta - A_2 \right)
- \frac{1}{z} \left( \beta - B_1 \!+\! 1 \right) 
	\left( \beta \!-\! B_2 \!+\! 1\right) \right\} \theta^{\,p-2} 
\;.\nonumber
\end{eqnarray}
The first condition is that there should exist a common factor for the 
last line in Eq.~(\ref{D0:2}), for example, $B_1 = A_1+1$, and 
consequently, $\beta=A_1.$\\[1ex]

\noindent 
{\bf Example IV: }
Let us consider the $\ep$ expansion of a hypergeometric function ${}_3F_2$
with the following set of parameters:
$
\omega(z) = 
{}_3F_{2}\left( \frac{r}{q}\!+\! a_1\ep, a_2 \ep, a_3 \ep; 
          1\!+\!\frac{r}{q}\!+\!\!b_1 \ep, 1\!-\!\frac{p}{q} 
	   \!+\!\!b_2 \ep;z\right)
$ with $a_1,a_2,a_3 \neq 0.$
Starting from the differential equation for this hypergeometric function,
\begin{equation}
\left[ 
z 
\left(\theta\!+\!\frac{r}{q}\!+\!a_1\ep\right)  
\left(\theta\!+\!a_2\ep\right)  \left(\theta\!+\!a_3\ep\right)  
\!-\! 
\theta 
\left(\theta\!+\! \frac{r}{q} \!+\!b_1\ep\right)
\left(\theta\!-\! \frac{p}{q} \!+\!b_2\ep\right)
\right] \omega(z) = 0 \;.
\label{de}
\end{equation}
the system of differential equations for the coefficient
functions $w_k(z)$ in its $\ep$ expansion
$
\omega(z) = 1 + \sum_{j=1}^\infty w_k(z) \ep^k 
$ 
is found to be
\begin{eqnarray}
&& 
\left[ 
(1\!-\!z) \frac{d}{dz} \!-\! \frac{1}{z} \frac{p}{q} 
\right] 
\left( \theta + \frac{r}{q} \right)
\theta 
\omega_m(z) 
= 
\left[ 
a_1+a_2+a_3
\!-\! 
\frac{b_1+b_2}{z} 
\right]
\left( \theta + \frac{r}{q} \right)
\theta 
\omega_{m-1}(z)
+ a_2 a_3 \frac{r}{q} \omega_{m-2}(z) 
\nonumber \\ && 
\qquad + a_1 a_2 a_3 \omega_{m-3}(z)
+ \delta_1 \theta \omega_{m-1}(z) 
+ \delta_2 \theta \omega_{m-2}(z) 
+ \frac{1}{z} \delta_3 \theta \omega_{m-1}(z) 
+ \frac{1}{z} \delta_4 \theta \omega_{m-2}(z) 
\;.
\label{3F2:1}
\end{eqnarray}
where $\theta=zd/dz$ and $\delta_j$ are constants.  After a redefinition 
\begin{eqnarray}
\left(
\omega_k(z), 
\theta \omega_k(z), 
\left( \theta + \frac{r}{q} \right) \theta \omega_k(z), 
\right) 
\to 
\left(
\omega_k(z), 
z^{-r/q} \sigma_k(z), 
\left(\frac{z}{z-1} \right)^{p/q} \phi_k(z)
\right) \;, 
\end{eqnarray}
we find that the function $h(z)$ defined by (\ref{h}) is 
$
h(z) = z^{(p+r)/q}(z-1)^{-p/q}.
$
By (\ref{R-relations}), a rational parametrization is possible when $p=-r$.


The methods described here can be extended to any multi-loop Horn-type 
hypergeometric function. 
Starting from the Pfaff form of the differential equation, 
\begin{eqnarray}
d \phi^{(i)}(\vec{z},\ep) = \sum_{k,j} P_{i,j,k}(\vec{z},\ep) \phi^{(j)}
	(\vec{z},\ep) dz_k \;,
\label{diff:pfaff:multi}
\end{eqnarray}
where $P_{i,j,k}(\vec{z},\ep)$ are rational functions, the system can be
transformed to triangular form and integrated.

Let us consider the $\ep$-expansion of the Appell hypergeometric function $F_3$,
which was analyzed in the context of photon box diagrams \cite{Davydychev:box}:
\begin{eqnarray}
&& 
F_3\left(\frac{p_1}{q} \!+\! a_1 \ep, \frac{p_2}{q}\!+\!a_2 \ep, 
         \frac{r_1}{q}\!+\!b_1\ep, \frac{r_2}{q}\!+\!b_2\ep, 
          1\!-\!\frac{p}{q}+c\ep; x,y
 \right)
\nonumber \\ && 
\qquad =
\sum_{m=0}^\infty
\sum_{n=0}^\infty
\frac{
\left( \frac{p_1}{q}\!+\!a_1\ep \right)_{m} 
\left( \frac{p_2}{q}\!+\!a_2\ep \right)_{n} 
\left( \frac{r_1}{q}\!+\!b_1\ep \right)_{m} 
\left( \frac{r_2}{q}\!+\!b_2\ep \right)_{n} 
}
{
\left( 1-\frac{p}{q}+c\ep \right)_{m+n} 
}
\frac{x^m}{m!}
\frac{y^n}{n!} \;.
\end{eqnarray}
Applying our methods, we find that the coefficients of 
$\ep$-expansion 
may be expressed in terms of Goncharov's polylogarithms when
$p_j r_j = 0$ for $ j = 1,2,$
and a rational parametrization should exist for the functions
\begin{eqnarray}
h_1(x) & = &  
(-1)^{s_1/q}
\left[ 
\frac{x^p}{(x-1)^{s_1+p}}
\right]^{1/q} 
\;, 
\quad 
h_2(x)  =   
(-1)^{s_2/q}
\left[ 
\frac{y^p}{(y-1)^{s_2+p}}
\right]^{1/q} 
\;, 
\\
\qquad H(x,y) & = &  
(-1)^{(s_1+s_2)/q}
\left[ 
\frac{x^{s_2+p}y^{s_1+p}}
     {(xy-x-y)^{s_1+s_2+p}}
\right]^{1/q} 
\;, 
\end{eqnarray}
where 
$s_j = p_j + r_j$ and $j = 1,2$.
%
%
As result of our analysis, we claim that in two cases (only), 
\begin{eqnarray}
&& 
F_3\left(I_1 + \frac{p_1}{q} + a_1 \ep, I_2+a_2 \ep, I_3+b_1\ep, I_4+b_2\ep, 
	I_5+\frac{p_1}{q}+c\ep; x,y \right) \;, 
\\ && 
F_3\left(I_1 + \frac{p_1}{q} + a_1 \ep, I_2+a_2 \ep, I_3+b_1\ep, I_4+b_2\ep, 
	I_5+c\ep; x,y \right) \;,
\label{result}
\end{eqnarray}
with integer values $I_j, p_1, q$,
the $\ep$ expansion $F_3$ can be expressed in terms of 
Goncharov polylogarithms.
We note that for $F_3$, the $\ep=0$ term should be a rational function.

\begin{acknowledgments}
S.A.Y.\ acknowledges the support of U.S. DOE grant DE-PS02-09ER09-01 and
The Citadel Foundation.  M.Yu.K.\ and B.A.K.\ were supported in part by the 
German Federal Ministry for Education and Research BMBF through Grant 
No.\ 05~HT6GUA, by the German Research Foundation DFG through the 
Collaborative Research Centre No.~676 {\it Particles, Strings and the 
Early Universe---The structure of Matter and Space Time}, 
and by the Helmholtz Association HGF through the Helmholtz
Alliance Ha~101 {\it Physics at the Terascale}. B.F.L.W.\ thanks 
Prof.\ I.\ Antoniadis for the support and kind hospitality of the CERN TH
Unit, and acknowledges partial support from U.S. DOE grant DOE-FG02-09ER41600.
\end{acknowledgments}

\bigskip 

\end{document}